\documentclass[runningheads]{llncs}

\usepackage{graphicx}
\usepackage{relsize}
\usepackage{fancyhdr}
\usepackage{subfigure}
\usepackage{lineno,hyperref}
\usepackage{bbm} 
\usepackage{xspace} 
\usepackage{booktabs}
\usepackage{tabulary}
\usepackage{tabularx}
\usepackage{color} 
\usepackage{epsfig} 
\usepackage{graphicx} 
\usepackage{float} 
\usepackage{enumerate} 
\usepackage{algpseudocode}
\usepackage{amsmath}
\usepackage{relsize}
\usepackage{amssymb} 
\usepackage{footnote} 
\usepackage{algorithm}
\usepackage{longtable}

\usepackage{blindtext} 

\newcolumntype{P}[1]{>{\centering\arraybackslash}p{#1}}

\newcommand{\ie}{\textit{i.e.,}\xspace}

\newcommand{\imagen}[4] 
{	
	\begin{figure}[H]
		\begin{center}
			\includegraphics[scale=#1]{#2}
		\end{center}
		\vspace{-3mm}
		\caption{#3}
		\label{#4}
	\end{figure}
}

\begin{document}

\title{On the Importance of Domain-specific Explanations in AI-based Cybersecurity Systems \\[2pt]
(Technical Report)}
\author{Jose N.\ Paredes$^1$, Juan Carlos L.\ Teze$^2$, \\
Gerardo I.\ Simari$^{1,3}$, Maria Vanina Martinez$^4$}
\titlerunning{Explanations in AI-based Cybersecurity Systems}
\authorrunning{J.N.\ Paredes, J.C.L.\ Teze, G.I.\ Simari, and  M.V.\ Martinez}
\institute{
$^1$Dept.\ of Comp.\ Sci.\ and Eng., Universidad Nacional del Sur (UNS) and \\
Institute for Computer Science and Engineering (ICIC UNS--CONICET), Argentina \\
$^2$Fac.\ de Cs.\ de la Adm., Universidad Nacional de Entre Ríos (UNER), Argentina \\
$^3$CIDSE, Arizona State University (ASU), USA \\
$^4$Dept.\ of Computer Science, Universidad de Buenos Aires (UBA), and \\
Institute for Computer Science Research (ICC UBA--CONICET), Argentina
}
\maketitle              
\begin{abstract}
With the availability of large datasets and ever-increasing computing power, there has been a growing use of data-driven artificial intelligence systems, which have shown their potential for successful application in diverse areas. However, many of these systems are not able to provide information about the rationale behind their decisions to their users.  Lack of understanding of such decisions can be a major drawback, especially in critical domains such as those related to cybersecurity. In light of this problem, in this paper we make three contributions: (i) proposal and discussion of desiderata for the explanation of outputs generated by AI-based cybersecurity systems; (ii) a comparative analysis of approaches  in the literature on Explainable Artificial Intelligence (XAI) under the lens of both our desiderata and further dimensions that are typically used for examining XAI approaches; and (iii) a general architecture that can serve as a roadmap for guiding research efforts towards the development of explainable AI-based cybersecurity systems---at its core, this roadmap proposes combinations of several research lines in a novel way towards tackling the unique challenges that arise in this context.
\end{abstract}

\section{Introduction and Related Work}
\label{intro}

With the advances and performance improvements in computing power, the enormous amount of data and information available in a wide range of platforms has become an important component in people's decision-making processes.
One important example of such platforms is social media, with products like Facebook and Twitter being widely used by people not only to keep in touch with friends, family, and acquaintances, but also to stay informed about current events.
Thus, people increasingly base their daily decisions on what they find on such platforms, making it a very useful tool but also a very vulnerable one---social media systems have been the vehicle for spreading false rumors with the potential for causing great damage~\cite{thuraisingham2020role}.
These situations have generated a growing interest in the development of tools to obtain information not only about users but also the contents of their posts.
This is only one example of a sensitive domain in which data analysis is playing an important role---cybersecurity is a problem of growing relevance that inherently impacts all facets of society.

The continuous increase in security problems has given rise to a growing demand for Artificial Intelligence (AI) approaches that have a vital impact in detecting both simple security risks as well as sophisticated cyber attacks.
There are many applications where AI can have a positive impact; however, there has been a general
agreement in the R\&D community that the tools being developed suffer from the limitation of {\em not being explainable}~\cite{gunning2017explainable}; there is therefore a clear need for understanding the decision-making process of these systems, which has become generally known as {\em Explainable AI} (XAI).
In this context, users of AI-based tools not only need to make effective use of them (for instance, to identify
incoming threats in enterprise networks or detect other kinds of malicious activity) but these efforts also need support that allows them to understand
how the system is reaching its conclusions, and allow them to interact with the system in a collaborative manner.
Adding explainability features to AI-based systems has thus been proposed as one way in which this need can be potentially mitigated.
Towards this end, we need to understand both what the users need for effective decision-making, what information the system has, and how it should make it available.
In this paper, our aim is to explore the design requirements that should be considered in building XAI systems in cybersecurity domains, which we argue have several unique and especially challenging features.

\smallskip
\noindent
{\bf XAI in Cybersecurity.}
The understanding of the decisions of automated systems, and the reasoning behind such decisions, play an important role in any human decision-making process. Despite its importance, and successful applications in both academic proof of concept developments and mature implementations in industry, XAI in cybersecurity has not received much attention.
We now briefly review the most salient works, all of which have been published in the past two years.

Perhaps the closest work in spirit to our approach is the proposal of explainable security presented in~\cite{vigano2020explainable}, which discusses several topics and future challenges to be explored.
However, their focus is on highlighting questions that should receive special attention such as who gives/receives the explanations, what constitutes a good explanation, when the explanation should be given, how to give explanations, where the explanations should be made available, and why the explanations are needed; our work can be seen as a continuation
of this proposal along some of the proposed lines.
Recently, there has been a surge of research on enhancing the explainability of deep neural networks;
in our domain of interest in particular, the work of~\cite{MahdavifarG20} proposes to do this
through a rule extraction process that allows to explain the causes of cyber threats.
In~\cite{SzczepanskiCPK20}, the authors present a concrete proposal for an Explainable Intrusion Detection System, together with an empirical evaluation of its prototype implementation that involves a combination of a Feed-Forward Artificial Neural Network and Decision Trees using a microaggregation heuristic.
Finally, other efforts have taken the approach of understanding both what information requirements the human needs for decision-making, as well as what information the AI components have. In~\cite{holder2021explainable} the authors present a use case for capturing these issues, seeking a guide for the development of future explainable systems that users can leverage.

There are several differences between our work and the approaches mentioned above.
Despite sharing the general motivation of studying explanations in the cybersecurity context, our work
not only focuses on providing guidelines for high-quality and timely explanations in the context of cybersecurity applications, but also on proposing a general architecture especially geared towards the generation of user-friendly explanations based on both knowledge- and data-driven AI building blocks. In this regard, we present various properties capturing those requirements that we assume an explainable cybersecurity system should fulfil, taking some first steps towards the development of hybrid AI-based systems that provide decision support via the leveraging of explainability as one of the central components in achieving this goal.

\smallskip
\noindent
{\bf Contributions.}
The main contributions of this work can be summarized as follows:
(i)~We propose and discuss desiderata for the explanation of outputs generated by AI-based cybersecurity systems;
(ii)~We carry out a comparative analysis of approaches  in the literature on Explainable Artificial Intelligence (XAI) under the lens of both our desiderata and further dimensions that are typically used for examining XAI approaches; and
(iii)~We propose a general architecture that can serve as a roadmap for guiding research efforts towards the development of explainable AI-based cybersecurity systems---at its core, this roadmap proposes combinations of several research lines in a novel way towards tackling the unique challenges that arise in this context.

The remainder of the paper is structured as follows. Section~\ref{sec.background} provides a brief overview of various problems related to cybersecurity, focusing on those most likely to require explanations.
In Section~\ref{sec.expDom}, we study several desirable properties that explainable cybersecurity systems should satisfy,
as well as tools for integrating AI-based approaches and explanability.
Finally, Sections~\ref{sec.futureChallengues} and~\ref{sec.conclusion} describe future challenges and conclusions, respectively. 
\section{A Brief Background on Problems in Cybersecurity}
\label{sec.background}

Cybersecurity is a vast, multidisciplinary field, and surveying it is clearly outside of the scope of this work.
Instead, we will analyze a set of typical ``basic'' problems that are often encountered in the fight against
malicious actors (who are also commonly referred to as ``attackers''). There are many types of cyber threats, some of which are newer ({\em e.g.}, those related to social platforms) and others
that have been around for decades ({\em e.g.}, those affecting enterprise systems)---we now provide an
overview of some key challenges faced in the cybersecurity arena:

\smallskip\noindent
\textit{Cyber Attribution:}
The ``attribution problem'' refers to determining who was responsible for a given  attack. This is a difficult problem mainly due to the effort required to find and process evidence that can lead to attributions, but also the fact that attackers can (and often do) plant false pieces of evidence in order to mislead security analysts.
Commonly employed attribution techniques include reverse-engineering, source tracking, and
honeypots, among others~\cite{nunes2018artificial}.
Clearly, these techniques usually involve the analysis of multiple sources.

\smallskip\noindent
\textit{Bots/Botnets:}
A {\em bot} is an autonomous software tool---examples range from simple scripts to sophisticated intelligent agents.
Large groups of bots connected and organized towards achieving a specific goal are typically called {\em botnets}.
The role of bots has in the last decade become increasingly prevalent in many (not necessarily malicious) cyber operations such as stock market trading, customer care, and social media platforms in general.
In contrast, malicious bots and botnets have been used to infect targets, stealing CPU cycles from infected hosts to carry out cryptocurrency mining activities, or disseminate misinformation.
Deciding whether or not an online user is being controlled by a person or a bot is essential to deterring or hindering malicious activities~\cite{ferrara2016rise}.	

\smallskip\noindent
\textit{Hacking Activities:}
Malicious cyber events are becoming more prevalent and difficult to detect, and account for enormous costs for organizations around the world.
Driven by the increasing scale and the occurrence of high-profile cybersecurity incidents, academic and industry researchers are showing a growing interest to improve the ability for predicting attackers' next moves.
In this direction, there are various open challenges to be addressed against malicious actors,
such as intention recognition or intrusion prediction~\cite{ahmed2017attack}, and enterprise-targeted attacks in
general~\cite{AlmukayniziMSNSS20}.

\smallskip\noindent
\textit{Open Source Chatter Analysis:}
Recent years have seen dramatic increases in malicious hacker activity on forums and marketplaces on the {\em darknet}
(the part of the Internet not indexed by search engines, where user identities are hidden from law enforcement).
Analyzing communications (or {\em chatter}) between different kinds of actors---such as buyers and sellers of zero-day exploits, or among hackers regarding the exploitability of certain vulnerabilites---has shown to be an effective way of obtaining open source intelligence (OSINT) that can counter the efforts of malicious activities both at the enterprise and consumer levels~\cite{nunes2016darknet}.

\smallskip\noindent
\textit{Misinformation:}
The deliberate spread of misleading or false information can cause varying levels
of damage in the form of {\em fake news}, propaganda, manipulated elections, and stock market bubbles.
Because of its capacity for exploiting vulnerabilities in social systems, fake news and misinformation in general can be classified as a cyber threat~\cite{caramancion2020exploration}.
This setting contains formidable challenges, such as identifying false or highly biased information, and the actors that contribute to its
distribution. Although different types of solutions have been proposed and progress is being made in this direction, current automated solutions are still unable to effectively mitigate the problem of misinformation on the Web, and it is currently considered to be one of the most challenging threats for users and content administrators.

\smallskip\noindent
\textit{Adversarial Deduplication:}
As a final example, consider situations where multiple virtual identities map to the same real-world entity.
The classic problem of {\em deduplication} in databases typically occurs due to involuntary situations, such as clerical errors or those arising from the use of automated data entry procedures.
On the other hand, {\em adversarial deduplication}~\cite{paredes2018leveraging} refers to the problem of detecting these
situations under the assumption that they occur as a result of malicious intent; many common practices fall under this
category, such as sock puppets, Sybil attacks, and malicious hacker activities in forums and marketplaces, as discussed above.
An interesting characteristic in the latter scenarios is the fact that actors typically seek to both remain anonymous, but also identifiable to the degree that their reputation remains intact within the community (for instance, as
the provider of high-quality malware).

\medskip

There are many other challenging areas related to cybersecurity that have been addressed with AI tools, and
there is clearly much work still to be done in addressing cybersecurity issues.
As a more general presentation of a set of open cybersecurity-related research questions and needs,
we propose a list of ten complex cybersecurity problems---presented in Figure~\ref{table.queries}---that motivate
the need for effective collaboration between human analysts and AI-based tools when identifying and preventing threat actions. The next section deals specifically with the problem of obtaining {\em explanations} from such tools, which
is a basic building block in the implementation of this ``human-in-the-loop'' (HITL) scheme.

\setlength{\extrarowheight}{2pt}
\begin{figure}[t]
\scriptsize
	\centering
	\begin{tabular}{|c| m{10.5cm} | }
		\hline
		\textbf{\textup{Problem} }& \hspace*{16em} \textbf{\textup{Description}}\\
		\hline
		P1 & How can emerging cybersecurity incidents be identified in a timely fashion, so as to provide early warning alerts?\\ \hline
		P2 & Which system features are potentially insecure?\\ \hline
		P3 & How can lightweight temporal models be leveraged in cyber attack prediction?\\ \hline
		P4 & How can indicators of risk be scalably and effectively aggregated to achieve threat intelligence in near real time?\\ \hline
		P5 & Is there software running on company C's computers at risk of being attacked?\\	\hline			
		P6 & How can cybersecurity-relevant discussions on social platforms be leveraged as indicators for
policing cybercriminal activities?\\ \hline
		P7 & Is there evidence that a viable exploit for vulnerability X is being pursued at this time?\\ \hline
		P8 & How does malicious content (such as fake news, malware, etc.) spread?\\ \hline
		P9 & How can malicious content be detected and addressed (labeled, blocked, etc.)?\\ \hline
		P10 & Can we identify fake profiles using data arising from social platform activity?\\ \hline
	\end{tabular}
	\caption{A sample of challenging problems that cybersecurity analysts typically address.}
	\label{table.queries}
\end{figure}

\section{Explainable Cybersecurity}
\label{sec.expDom}

As we began to argue in the previous section, the ability to understand and trust the decisions suggested by AI-based systems is relevant and necessary for users to be able to effectively operate in a collaborative manner with these
tools. This is especially so when applications are related to domains where decisions can have an important impact, such
as those related to the different areas of cybersecurity.
With this motivation, in this section we propose a set of properties for the study of explanation generation processes
in such domains.

\subsection{Desiderata for Explanations in Cybersecurity}

Explainability is not a new concept in AI; in order to characterize what is a good explanation, the authors of~\cite{swartout1993explanation} defined basic desiderata that include {\it fidelity}, {\it understandability},
{\it sufficiency}, {\it low construction overhead}, and {\it efficiency}.
We now focus on extending these desiderata by incorporating aspects that are especially relevant in cybersecurity applications. This is particularly challenging due to several characteristics; in particular, it typically involves several different stakeholders and is multi-faceted in nature since it requires reasoning about
different security components such as mechanisms, policies, threat model, and attack surface,
as well as concrete attacks, vulnerabilities, and countermeasures.

Despite the importance of explanations in establishing user trust in systems,
there is no consensus in the AI community regarding what constitutes  a \textit{useful explanation}.
Even though explainable AI (XAI) has been a very active research topic in the last few years, the topic of XAI in
cybersecurity has received little attention.
Motivated by this, in this section we will introduce a set of desirable properties that explanations should have in
cybersecurity tools. To facilitate this task, we will first perform a more detailed analysis of relevant characteristics of cybersecurity domains that will be used in the rest of the paper.

\smallskip\noindent
\textit{(i) Dynamism}:
User behavior and available data in this type of setting are inherently dynamic;
in particular, the result of reasoning carried out in the past is susceptible to constant updates.

\smallskip\noindent
\textit{(ii) Time constraints}: Security problems often have potentially catastrophic consequences for users, organizations, and their clients.
Thus, users and analysts in HITL systems operate in time-sensitive conditions ({\em i.e.}, identifying and removing erroneous information sources, or simply changing a password).

\smallskip\noindent
\textit{(iii) Variety of actors}:
What makes cybersecurity unique is that it is inherently a combination of both human and highly computerized efforts.
Human participants occupy different roles (\ie the range between novice and expert users).
Examples include basic end users (typical consumers), power users (specialized operators), as well security analysts
that need to have access to all available details. Notably, malicious actors are also typically on this spectrum, ranging from novices playing with Metasploit to professional penetration testers or malicious state hackers.

\smallskip\noindent
\textit{(iv) Adversarial settings}:
There are two main kinds of entities interacting in cybersecurity domains:
{\em stakeholders} and {\em adversaries}. The first ones are comprised of system users and other interested parties,
whereas adversaries or attackers interfere with the stakeholders' goals.
In some cases, a stakeholder can become an adversary with respect to another system; for instance,
one company may be protecting its assets while spying on other companies.

\smallskip\noindent
\textit{(v) Heterogeneity}: Information typically arrives from multiple sources, and managing information flow is crucial in addressing many cybersecurity issues. This involves problems such as representing, modeling, and fusing  interconnected heterogeneous data and information derived from different sources.

\smallskip\noindent
\textit{(vi) Uncertainty}: Real-world data and information generation processes typically involve uncertainty---only
the simplest situations avoid this issue. Uncertainty has two extremes: overspecification (or inconsistency) and underspecification (or incompleteness). The temporal and dynamic nature of our setting
involves new information arriving periodically, so the picture is in constant evolution.

\smallskip\noindent
\textit{(vii) Socio-techno-political aspects}:
The role of data in a wide range of platforms has in the last decade had large influence on many people's
decision-making processes related to politics, economics, entertainment, and many others.
Clearly, the increasingly central role of users in these environments creates opportunities for malicious actors,
and the landscape of cybersecurity issues is expanding~\cite{bella2015service}.

\medskip
\noindent
{\bf The Desiderata.}
Apart from obtaining the answers to a specific cybersecurity query, system users may need informative explanations regarding such answers.
With this motivation, we study what constitutes a good explanation within the context of cybersecurity, focusing
on the inherent properties discussed above.
We first propose a set of desired characteristics and capacities that explanation methods should have;
we will come back to them later when analyzing the applicability to cybersecurity of proposals from the general
XAI literature.

\smallskip\noindent
(D1) \textit{Temporality-awareness}:
Attackers change their strategies dynamically depending on measures taken against them, so security systems need to
adapt to constantly changing environments.
This synergy highlights new demands that explanations should capture---an explanation revealing details of internal data requested at time $T_{i}$ possibly changes at time $T_{i + 1}$. Given this consideration, temporal aspects such as specific time points or intervals should be explored by explanation-generation methods.

\smallskip\noindent
(D2) \textit{Abnormality-awareness}:
The tools specially designed to address tasks related to the detection of malicious behavior typically tend to focus more on events or observations that are considered to be unexpected or unusual.
When an abnormality is detected, a report may be required with explanatory information for investigating the report in detail. Abnormality clearly plays a central role in explanations in the cybersecurity context, and abnormal events could serve as a trigger for explanations. However, in the case of having characterizations of (ab)normal situations, we could take advantage of them and provide {\it abnormality-centered} explanations related to cybersecurity problems by identifying what (ab)normal situations have occurred.
	
\smallskip\noindent
(D3) \textit{Human-centeredness}:
Given the increasing complexity of the threats and sensitive user data they affect, security systems must adopt approaches capable of providing users with tools that increase their transparency and trust in the system's outputs.
The benefits afforded by explanations only fully come to bear when human users are able to {\em understand and interact} with them.
To develop effective explainable methods to support this need, it is key to establish how these models can be easily integrated into human-in-the-loop decision systems.

\smallskip\noindent
(D4) \textit{Inconsistency-awareness}:
Explanation quality can be affected by inconsistencies in the underlying data; thus, the ability to identify them
can be of great value.
Note that inconsistencies can appear in: (i) basic data, (ii) answers/outputs, or (iii) both data and processed outputs.

\smallskip\noindent
(D5) \textit{Uncertainty-awareness}:
As discussed, cybersecurity domains involve features that lead to potentially high degrees of uncertainty. Thus, handling uncertainty in explanation generation methods is especially useful for situations where it is generally impossible to treat data and inferences with complete certainty.

\medskip

Next, we examine several XAI approaches proposed in the literature in light of the features and desiderata discussed above.

\subsection{A Survey of Relevant XAI Approaches in the Literature}

We now carry out a comparative analysis of XAI approaches by considering the desiderata for cybersecurity explanations and the following further dimensions:

\begin{itemize}
	\item \textbf{Method:} Two types of methods to provide explanations;
{\em local} (L) ones give explanations via individual instances or groups of nearby instances, and
{\em global}~(G) ones that describe the behavior of models as a whole.

\item \textbf{User:} Two kinds of stakeholders: {\em novice} (N) and {\em expert} (E).
	
\item \textbf{Model:} Approaches to provide explanations can be classified as
(i) {\it Model-agnostic} (A), which apply to any model (based only on inputs and outputs), and
(ii) {\it Model-specific} (S), which are restricted to a specific model.

\item \textbf{Contrastiveness:} Refers to approaches where explanations are centered on a contrastive strategy; {\em e.g.}, counterfactuals are a particular case of this.

\item \textbf{Uncertainty:} Refers to approaches that incorporate uncertainty in their explanations, either in a quantitative or qualitative manner.
\end{itemize}

\begin{figure}[t]
\renewcommand{\arraystretch}{1.2}
\centering
\scriptsize
\begin{tabular}{| l | c | c | c | c | c |}
			\hline
			\textbf{Name} & \ \  \textbf{Method}\ \  & \ \ \ \ \textbf{User}\ \ \ \  & \ \ \textbf{Model} \ \ & \textbf{Contrastive} & \textbf{Uncertainty}\\
			\hline
			Explanation Vectors (2010)~\cite{Baehrens2010explain} & L & E & A & No & Yes\\
			\hline
			LIME (2016)~\cite{Ribeiro2016should} & L & E & A & No & Yes\\
			\hline
			SHAP (2017)~\cite{LundbergL17} & L & E & A & No & Yes\\
			\hline
			Grad-CAM (2017)~\cite{Selvaraju2017grad} & L & N & S & Yes & Yes\\
			\hline
			ANCHORS (2018)~\cite{Ribeiro2018anchors} & L & N & A & No & Yes\\
			\hline
			BEEF (2019)~\cite{Grover2019beef} & G & N & A & Yes & Yes\\
			\hline
			I-GOS (2019)~\cite{Qi2019visualizing} & L & N & S & No & Yes\\
			\hline
			Justified Counterfactual (2019)~\cite{Laugel2019dangers} & L & E & A & Yes & No\\
			\hline
			Prototype and Criticism (2016)~\cite{KimKK16} & G/L & E & A & Yes & No\\
			\hline
			DeNNeS (2020)~\cite{MahdavifarG20} & G/L* & E & S & No & No\\
			\hline
			Hybrid Oracle-Explainer IDS~\cite{SzczepanskiCPK20} & L & N & A & No & Yes\\
			\hline
			TreeExplainer (2020)~\cite{Lundberg2020local} & L & E & S & No & Yes\\
			\hline
			AIP (2021)~\cite{Yang2021generative} & L & N & A & Yes & No\\
			\hline
			DeLP3E (2016)~\cite{ShakarianSMPPFA16} & G/L* & E & S & No & Yes\\
			\hline
			EMDM (2019)~\cite{zhong2019explainable} & G & N & S & Yes & No\\
			\hline
			A-I model (2018)~\cite{Rago2018argumentation} & G & E & S & No & No\\
			\hline
			ADSS with CPref-rules (2021)~\cite{BuronTG21} & G & E & S & No & No\\ \hline
\end{tabular}
\caption{XAI approaches and their main characteristics concerning our desiderata and other typical dimensions.
Symbol ``$*$'' in Method refers to cases where it is not clear if explanations can be based on global or local methods.}
			\label{tab:exp_sys}
\end{figure}

Our main findings are summarized in Figure~\ref{tab:exp_sys}---note that D1--D4 do not appear because
these features are not present in any of them.
We now discuss some highlights of our analysis.

\smallskip
\noindent
{\em Global vs.\ Local.}
Local methods are the most common, which is perhaps due to the fact that explanations are generally required for the most complex AI-systems where many entities and features/attributes have to be analyzed, which leads to a more challenging problem for global methods.

\smallskip
\noindent
{\em Expert vs.\ Novice.} Explanations requiring expert knowledge are more common, perhaps because making technical knowledge available to lay users can be challenging. However, considering both kinds of users is crucial because of the multiplicity of actors involved in cybersecurity domains, as discussed above.

\smallskip
\noindent
{\em Specific vs.\ Agnostic.} There exist many works both with and without assumptions about the reasoning mechanism of the system. These two approaches are in opposition; generally, the former can provide good explanations by taking advantage of available knowledge about internal structure of the model but it clearly has a more limited scope.
Instead, the latter has greater flexibility but the computational cost can become high as the number of available
features/attributes increases, and the selection of the instances/entities to build the explanation model can become a difficult barrier to overcome.

\smallskip
\noindent
{\em Contrastiveness.}
This is not a popular feature in the literature, likely because it is not an easy task to produce appropriate
comparisons.
Counterfactual approaches (such as~\cite{Laugel2019dangers,Yang2021generative,Selvaraju2017grad}) are a particular case where the contrast is based on synthetic situations (values of features/attributes); another example is that of
{\em balanced} explanations, which appear in~\cite{Grover2019beef}.

\smallskip
\noindent
To conclude, we focus on two more general features:

\smallskip
\noindent
{\bf KR-based.}
Traditionally, logic-based reasoning has been studied as more closely related to the way humans reason, and
thus the path to good explanations is believed to be shorter---many works in KR\&R have considered that
explainability is a feature that ``comes built in'', though this is not necessarily so simple.
For instance, even if sets of rules of the form {\it if body then head} are similar to those used in human decision-making, the semantics of different sets can differ, or the set can be very large, which greatly increases the complexity of the task of deriving explanations.
Many KR-based works, such as~\cite{zhong2019explainable,Rago2018argumentation,BuronTG21} (included
in Figure~\ref{tab:exp_sys}), address the explanation problem in AI-based systems that are not centered in cybersecurity domains. These approaches are in general specific and require expert users, but they consider that merely implementing symbolic reasoning is not necessarily enough for providing explanations; hence, different improvements are provided in order to reach a greater understanding about the reasoning process ({\em e.g.}, using appropriate views with several detailed levels for this purpose).

\smallskip
\noindent
{\bf Cybersecurity-friendliness.} As can be seen in Figure~\ref{tab:exp_sys}, none of the approaches covers all properties that we consider important in cybersecurity domains.
In fact, we recall that (i) none of the approaches consider D1--D4, and (ii) the number of approaches with and without the ability to handle uncertainty (D5) is balanced. We consider this to be a clear indicator of the lack of research in the area of explainable AI-System specifically for cybersecurity domains.

\smallskip
\medskip
From the literature analysis, we can conclude that even though there are many tools being developed for XAI, advances are not particularly centered on cybersecurity domains. Furthermore, the growing popularity of data-driven (ML) approaches has also been mirrored in the attention that explanations have received in this subarea of AI.
Although it is clear that explainable data-driven models is a relevant topic that can yield great benefits, we consider that further studies are also necessary in broader AI, seeking solutions that meet our desiderata for explanations in cybersecurity. This working hypothesis is linked with our view about how challenging issues such as cybersecurity problems can be effectively addressed via a combination of data-driven and KR-based models.

\section{Towards an Architecture for XAI in Cybersecurity}
\label{sec.futureChallengues}

Based on the proposed domain properties, desiderata, and literature survey,
we now introduce a general framework, whose architecture is illustrated in Figure~\ref{fig:exp_cyb_arch}, for guiding the implementation of explainable cybersecurity software systems.
The proposed framework, which generalizes the system developed for cybersecurity tasks in social platorms in~\cite{badbot}, consists of six main components:

\begin{figure}[t]
	\centering
	\includegraphics[width=0.8\textwidth]{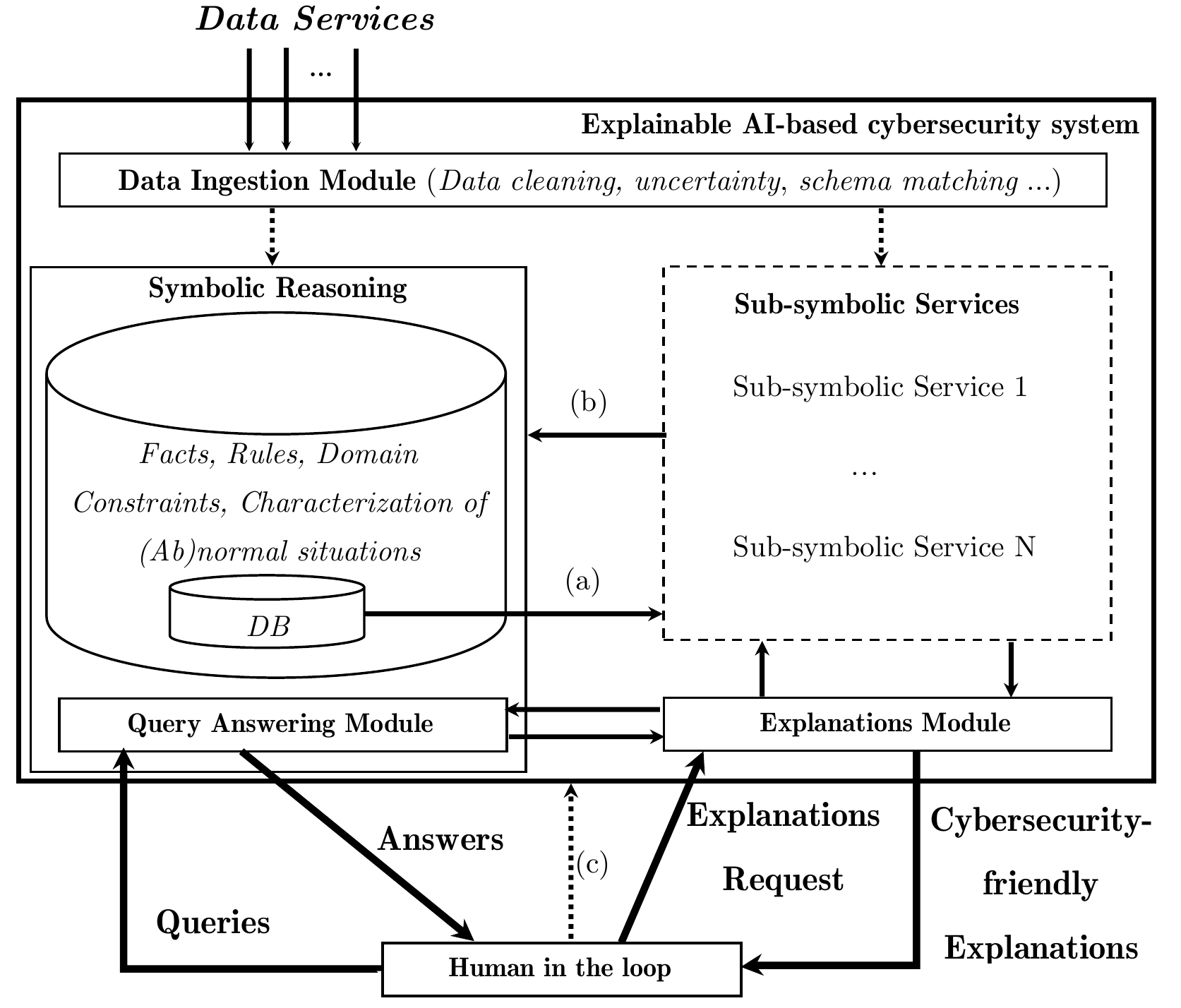}
	\caption{Generic architecture for XAI-based cybersecurity systems;
(a)~Sub-symbolic service invoked with data in DB, (b)~results used in rules, (c)~modifications by humans.}
	\label{fig:exp_cyb_arch}
\end{figure}

\smallskip\noindent
{\bf Data Services:}
A wide range of data sources may be required in order to effectively address the problems of interest.
For example, we could require data about vulnerabilities from the National Vulnerability Database\footnote{https://nvd.nist.gov/} and MITRE CVE\footnote{https://cve.mitre.org/},
types of common weakness in hardware and software from MITRE CWE\footnote{https://cwe.mitre.org/}, common tactics and techniques used by adversaries from MITRE ATT\&CK\footnote{https://attack.mitre.org/}, posts and relations from vlearnet ({\em e.g.}, tweets by cybersecurity experts) and darknet social platforms, among others.

\smallskip\noindent
{\bf Data Ingestion Module:}
Several hurdles arise from the integration of heterogeneous data sources, such as incompatible name spaces, inconsistency, uncertainty, structured, semi-structured, and/or unstructured data, different formats, among others.
This module is responsible for carrying out data preparation tasks such as data cleaning, data analytics, schema matching, inconsistency and incompleteness handling, etc., and also plays an important role for other kinds of more general issues such as trust and uncertainty management.

\smallskip\noindent
{\bf Sub-symbolic Services:}
Certain tasks are handled very effectively by data-driven models, so the system needs to provide a library
of available services, how they are implemented (tools used), what data are required, and how they can be updated.
Limiting the scope of such models allows to isolate application scenarios for each service in order to deploy them faster, tune different parameters, replace one implementation with another, and build more faithful explanations.

\smallskip\noindent
{\bf Symbolic Reasoning:}
We propose that more general problems are better addressed by high-level reasoning, which feeds from data pre-processed by the {\it Data Ingestion Module}, wile basic symbols are generated using {\it Sub-symbolic Services}.
Thus, a rule-based system can be appropriate to handle this high-level task since defined rules allow to easily
(i)~combine data and low-level knowledge,
(ii)~update and replace high-level knowledge,
(iii)~answer queries based on well-defined reasoning mechanisms ({\em cf.} the {\it Query Answering Module}), and
(iv)~provide explanations taking advantage of the access to well-defined inference processes.

\smallskip\noindent
$\bullet$ {\em Domain/Integrity constraints:}
Constraints allow us to detect inconsistency issues (D4). Thus, we could solve inconsistencies by updating and changing data before answering queries, or we could answer queries directly using available data but via inconsistency-tolerant semantics.

\smallskip\noindent
$\bullet$ {\em Characterization of (ab)normal situations:}
As mentioned before, cybersecurity problems typically involve abnormal events (D2);
therefore, it could be quite helpful to have conceptualizations about this kind of behaviors both to detect them and to explain results.
Similarly to the domain constraints, this characterization can be useful for both {\it Symbolic Reasoning} and {\it Explanations} modules.

\smallskip\noindent
{\bf Explanations Module:}
Explanations are associated with query answers and can be derived by accessing the {\it Symbolic Reasoning} module and the {\it Sub-symbolic Services}.
Systems with definitions of domain constraints and characterization of (ab)normal situations can be beneficial
to attaining desiderata D2 and D4.

\smallskip\noindent
{\bf Human in the Loop:}
This component addresses desideratum D3, which seeks iteratively administered feedback from human users who
can help improve system performance based on {\it queries}, {\it answers}, {\it explanation requests}, and
feedback in the form of explanation scoring, ranking of data sources by usefulness, etc.

\subsection{Example Applications}

We now briefly discuss two applications based on general problems included in Figure~\ref{table.queries}.
For problem P1; associated queries posed to the system could include:

\smallskip
\noindent
{\bf Q1.1}: {\em What are the list of all suspicious login attempts in the last 12 hours?};

\smallskip
\noindent
\noindent
{\bf Q1.2}: {\em What users have been involved in potentially malicious activities?}; and

\smallskip
\noindent
\noindent
{\bf Q1.3}: {\em What software components must be updated with priority given the system's
exposure to unpatched vulnerabilities?}

\smallskip
\noindent
As another example, for problem P9 the following are possible associated queries:

\smallskip
\noindent
{\bf Q9.1}: {\em What posts correspond to attempts at disseminating misinformation?};

\smallskip
\noindent
{\bf Q9.2}: {\em Which users are suspected of having multiple accounts on the system to avoid being
detected in malicious activities?}; and

\smallskip
\noindent
{\bf Q9.3}: {\em What is the projected effect of labeling specific posts on the overall reach of
fake news articles?}

\smallskip
In each case, our proposed model can be used to identify which aspects of the query can be tackled
with rule-based systems, where off-the-shelf ML-based tools can be effectively leveraged, and
how explanations can be built for each specific stakeholder.
An example of such developments is our recent work on detecting malicious behavior in social
platforms~\cite{badbot}, though explainability has not yet been addressed in that research line.

\subsection{Future Challenges and Research Opportunities}

There are several formidable hurdles faced by the cybersecurity community that must be addressed before obtaining
fully working solutions. The following is a high-level roadmap for research and development in this line of work.

\medskip\noindent
{\em Stream reasoning}: This research area aims at studying how to perform logical inference processes over data that is highly dynamic and cannot be stored at once; it has developed within several areas of study such as
Data Stream Management Systems (in the Databases community), Complex Event Processing (in Even-based Systems),
and the Semantic Web (in AI)~\cite{dell2017stream}.
One element usually found in stream reasoners is the presence of formal approaches that possibly combine stream logic reasoning with other techniques, such as machine learning and probabilistic reasoning.
To address desideratum D1, temporal stream reasoning could supply a suitable alternative solution in this direction.

\medskip\noindent
{\em Human-computer interaction (HCI)}:
This research area seeks to find knowledge about how to design and use {\it interfaces} with the intention of effectively creating a bridge between users and computers. It is clear that its scope extends far beyond AI, but this is an essential research line for {\it human in the loop} systems (which we consider to be central to cybersecurity) where AI models are leveraged by obtaining knowledge from interactions with users~\cite{Zanzotto19,NashedB18}.
In Figure~\ref{fig:exp_cyb_arch}, arrows to and from the ``{\it Human in the Loop}'' box indicate interactions between users and the system; hence, it is necessary to research which are the best ways humans can pose queries, be informed about answers, request explanations, and transfer knowledge to and from the cybersecurity system.

\medskip\noindent
{\em Natural language processing (NLP)}:
This field arises from three different disciplines: linguistics, computer science, and AI.
It aims to study the interactions between computers and human language, which can involve speech recognition, natural language understanding, and natural language generation.
The capabilities can range from accurately extracting information and insights contained in unstructured data to categorizing and organizing the data units.
In the context of Figure~\ref{fig:exp_cyb_arch}, research in this area can be beneficial to the implementation of {\it Sub-symbolic Services}, but also collaborations with the HCI community can improve the interactions between humans and the system (as in, for instance, \cite{MaidenZBBNTAE18}).

\medskip\noindent
{\em Computational Argumentation}:
This area studies how reasons for and against conclusions can be leveraged in automated reasoning processes
that are inspired in the way (rational) humans argue.
To provide explanations, conclusions can be accompanied by structures that are meaningful approximations of the whole reasoning process. Following Figure~\ref{fig:exp_cyb_arch}, works in this area can be useful for {\it Symbolic Reasoning} or the {\it Explanation Module} by extending studies such as~\cite{BuronTG21} or~\cite{Rago2018argumentation}. We can also take advantage of combining data-driven explanation tools ({\em cf.} Figure~\ref{tab:exp_sys}) and this area in order to achieve augmented explanations about the answers coming from the {\it Sub-symbolic services}.

\medskip\noindent
{\em Ontological Reasoning}:
This area arises at the intersection between Databases and AI. Ontologies are high-level representations about a particular domain, which are described using a rich semantic formalism built under Knowledge Representation and Reasoning principles, and typically techniques based on computational logic.
These representations structure and limit what is relevant to an application domain; many reasoning tasks can be formalized via specialized query answering mechanisms, and answers can be explained via a meaningful subset of this process.
Findings in this area can enhance the {\it Symbolic Reasoning} module---as in the previous paragraph,
explanations for answers provided by the {\it Sub-symbolic services} can be combined with the aim of achieving augmented explanations based on data-driven models, as proposed recently in~\cite{ConfalonieriWBM21}.

\section{Conclusions}\label{sec.conclusion}

Explainable Artificial Intelligence has witnessed unprecedented growth in recent years.
Even though many approaches for generating explanations have been advanced to support human decision-making processes in diverse areas, achieving explainability for cybersecurity systems has only begun to be explored in the literature.
Considering the complexity and dynamic nature of cyber threats, it has become increasingly desirable to understand the logic behind the outputs of these systems;
in this work, we have taken a significant step in investigating how software can be developed with explanatory power in this challenging arena.

Our contributions include (i) the proposal of a set of desiderata to guide the development of such tools,
(ii) a detailed comparative analysis between different approaches in XAI based on our desiderata and a set of additional dimensions that are commonly taken into consideration in the literature, and
(iii) a framework proposed as a general model for guiding the implementation of explainable software tools in cybersecurity domains that integrates several research lines in a novel way.
A valuable feature of this model is its capacity for addressing multiple problems at once.

Future work involves further developing this research line by studying new developments to tackle the limitations and challenges that we discussed in Sections~\ref{sec.expDom} and~\ref{sec.futureChallengues}.
We are currently working on fully formalizing our proposed model, and extending the implementation already developed
in~\cite{badbot} with explainability features that are capable of addressing the desiderata proposed in this work.

\bibliographystyle{splncs04}
\bibliography{bibliografia}
\end{document}